\documentclass[aps,groupedaddress,12pt]{revtex4-1}
\usepackage[utf8x]{inputenc}
\usepackage[T1]{fontenc}
\usepackage[pdftex]{hyperref,graphicx}	
\hypersetup{colorlinks,%
  citecolor=blue,%
  filecolor=green,%
  linkcolor=blue,%
  urlcolor=blue,%
 }
\usepackage{pdfpages}

\begin{document}

\title{Chimera States in Mechanical Oscillator Networks}

\email[Corresponding authors:]{$\;$erik.martens@ds.mpg.de, shashi@princeton.edu}
\altaffiliation{$^\dag$Equal author contribution}

\author{
Erik A. Martens$^{1,2,\ast,\dag}$,
Shashi Thutupalli$^{1,3,\ast,\dag}$,
Antoine Fourrière$^1$,
Oskar Hallatschek$^{1,4}$}

\affiliation{
$^1$Max Planck Institute for Dynamics and Self-Organization, 37077 Göttingen, Germany\\
$^2$Technical University of Denmark, 2800 Kgs. Lyngby, Denmark\\
$^3$Dept. of Physics and Dept. of Mechanical and Industrial Engineering, Princeton University, Princeton, NJ 08544 USA\\
$^4$Dept. of Physics, University of California, Berkeley, CA 94720, USA
}

\date{\today}

\begin{abstract}
The synchronization of coupled oscillators is a fascinating manifestation of self-organization that nature employs to orchestrate essential processes of life, such as the beating of the heart. While it was long thought that synchrony or disorder were mutually exclusive steady states for a network of identical oscillators, numerous theoretical studies in recent years revealed the intriguing possibility of `chimera states', in which the symmetry of the oscillator population is broken into a synchronous and an asynchronous part.
However, a striking lack of empirical evidence raises the question of whether chimeras are indeed characteristic to natural systems. This calls for a palpable realization of chimera states without any fine-tuning, from which  physical mechanisms underlying their emergence can be uncovered. 
Here, we devise a simple experiment with mechanical oscillators coupled in a hierarchical network to show that chimeras emerge naturally from a competition between two antagonistic synchronization patterns. We identify a wide spectrum of complex states, encompassing and extending the set of previously described chimeras. Our mathematical model shows that the self-organization observed in our experiments is controlled by elementary dynamical equations from mechanics that are ubiquitous in many natural and technological systems. The symmetry breaking mechanism revealed by our experiments may thus be prevalent in systems exhibiting  collective behaviour, such as power grids, opto-mechanical crystals or cells communicating via quorum sensing in microbial populations.  
\end{abstract}





\maketitle

Christiaan Huygens observed in 1665 that two pendulum clocks, suspended on a beam, always ended up swinging in exact anti-phase motion~\cite{Huygens1967ab} regardless of the pendula's initial displacements. He explained this self-emergent synchronization as resulting from the coupling between the clocks, mediated by vibrations traveling across the beam. Huygens' serendipitous discovery has  inspired many studies since, to establish that self-emergent synchronization is a central process to a spectacular variety of natural systems, including the beating of the heart~\cite{Michaels1987}, flashing fireflies~\cite{Buck1968}, pedestrians on the bridge locking their gait~\cite{Strogatz2005}, circadian clocks in the brain~\cite{Liu1997}, superconducting Josephson junctions~\cite{Wiesenfeld1998},  chemical oscillations~\cite{Kiss2002,Taylor2009}, metabolic oscillations in yeast cells~\cite{Dano1999} and life-cycles of phytoplankton~\cite{Massie2010}.

Ten years ago, the dichotomy between synchrony and disorder was challenged by a theoretical study revealing that a population of identical coupled oscillators can attain a state where one part synchronizes and the other oscillates incoherently~\cite{Motter2010a,Kuramoto2002,Abrams2004,Abrams2008,Montbrio2004,Omelchenko2008,Pikovsky2008,Bordyugov2010,Martens2010bistable,Martensswc2010,Laing2009,Laing2012a,Olmi2010}. These `chimera states'~\cite{Abrams2004} emerge when the oscillators are coupled non-locally, i.e., the coupling strength decays with distance between oscillators -- a realistic scenario in many situations including Josephson junction arrays~\cite{Phillips1993} or ocular dominance stripes~\cite{Swindale1980}. 
Chimera states are counterintuitive because they occur even when units are identical and coupled symmetrically; however, with local or global coupling, identical oscillators either synchronize or oscillate incoherently, but never do both simultaneously. 

Since their discovery, numerous analytical studies~\cite{Abrams2004,Abrams2008,Omelchenko2008,Pikovsky2008,Bordyugov2010}, involving different network topologies~\cite{Abrams2008,Martens2010bistable,Martensswc2010}, and various sources of random perturbations~\cite{Laing2009,Laing2012a} establish chimeras as a robust theoretical concept and suggest that they exist in complex systems in nature with non-local interactions. 
Yet, \emph{experimental} evidence for chimeras has been particularly sparse so far and has only been achieved recently via a computer-controlled feedback~\cite{Tinsley2012,Hagerstrom2012}. This raises the question whether chimeras can only be produced under very special conditions or whether they arise  via generic physical mechanisms. Uncovering such physical mechanisms requires analytically tractable experiments with direct analogues to natural systems.

\begin{figure*}[htp!]
 \centering
 \includegraphics[width=0.85\textwidth]{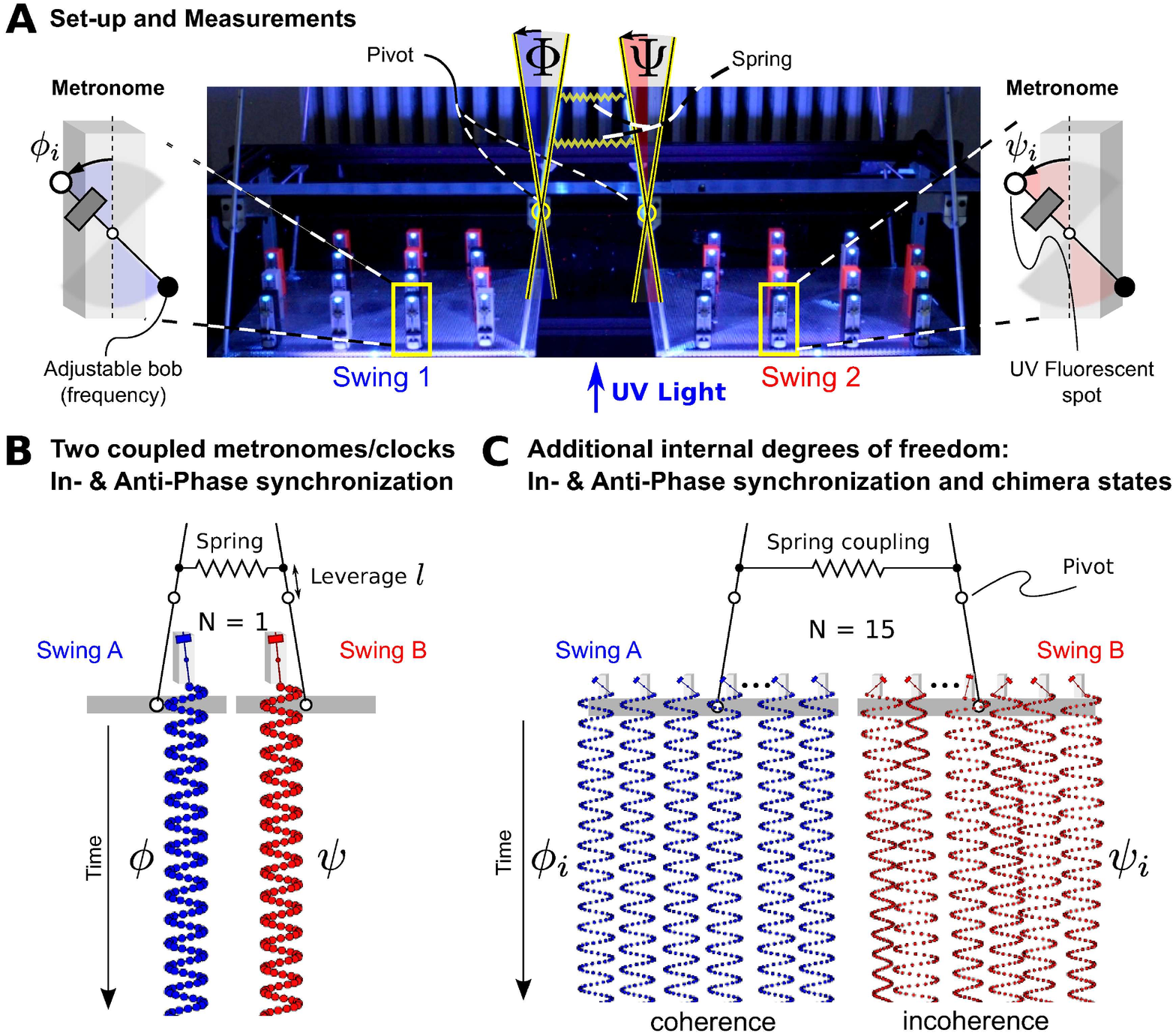}
 \caption{Experimental setup and measurements. Two swings are loaded with $N$ metronomes each and coupled with adjustable springs. 
 ({\bf A}) Swing and metronome displacements 
 are measured by digital tracking of UV fluorescent spots placed on pendula and swings. 
 ({\bf B}) $N=1:$ metronomes synchronize in anti- or in-phase motion. ({\bf C}) $N=15:$ symmetry breaking chimera states with one metronome population synchronized and the other desynchronized, or vice versa. $\phi_i$ and $\psi_i$ are the pendula's displacement angles on the left and right swing, respectively.}
 \label{figure1}
\end{figure*}

Our mechanical experiment 
shows that chimera states emerge naturally without the need to fine-tune interactions. 
We implement  the simplest form of nonlocal coupling that can be achieved using a hierarchical network with two subpopulations~\cite{Abrams2008,Montbrio2004}: within each subpopulation, oscillators are coupled strongly, while the coupling strength between the two subpopulations is weaker. 
We place $N$ identical metronomes~\cite{Pantaleone2002} with nominal beating frequency $f$ on two swings, which can move freely in a plane (Fig.~\ref{figure1}). 
Oscillators within one population are coupled strongly by the motion of the swing onto which the metronomes are attached. 
As $f$ is increased, more momentum is transferred to the swing, effectively leading to a stronger coupling among the metronomes.  
A single swing follows a phase transition from a disordered to a synchronized state as the coupling within the population increases~\cite{Pantaleone2002,Ulrichs2009}. This mimics the synchronization of pedestrians' gait on the Millennium Bridge~\cite{Strogatz2005} wobbling under the pedestrians' feet. In our setup, emergent synchronization can be perceived both aurally (unison ticking) and visually (coherent motion of pendula). Finally, the weaker coupling between the two swings is achieved by tunable steel springs with an effective strength $\kappa$.

\section{Results}

For nonzero spring coupling, $\kappa>0$, we observe a broad range of parameters in which chimeras (Fig.~\ref{figure1}C and Movie S1) and further partially synchronized states emerge. To quantitatively explore this complex behavior, we measure the metronomes' oscillation phase $\theta_k$, their average frequencies $\bar{\omega_k}$ and the complex order parameter, $Z_p(t)=N^{-1}\sum^N_{k=1}e^{i[ \theta_k^{(p)}(t)- \bar{\theta}_\textrm{\tiny syn}(t)]}$, where $p=1,2$ denotes the left or right population and $\bar{\theta}_\textrm{\tiny syn}$ is the average phase of the synchronous population ($|Z|$ quantifies the degree of synchronization: $|Z|\approx 0$ for incoherent and  $|Z|\approx1$ for synchronous motion).

To investigate where chimeras emerge in parameter space, 
we have systematically varied the effective spring coupling, $\kappa$, and the nominal metronome frequency, $f$, while ensuring that the metronomes on uncoupled swings synchronize. The long-term behavior of the system~\cite{Kuramoto2002,Abrams2004} is studied by preparing the experiments with several initial conditions (see SI Appendix): (i) both populations are desynchronized (DD) or (ii) one population is synchronous and the other desynchronized (SD and DS, respectively).
\begin{figure*}[htp!]
 \centering
 \includegraphics[width=0.85\textwidth]{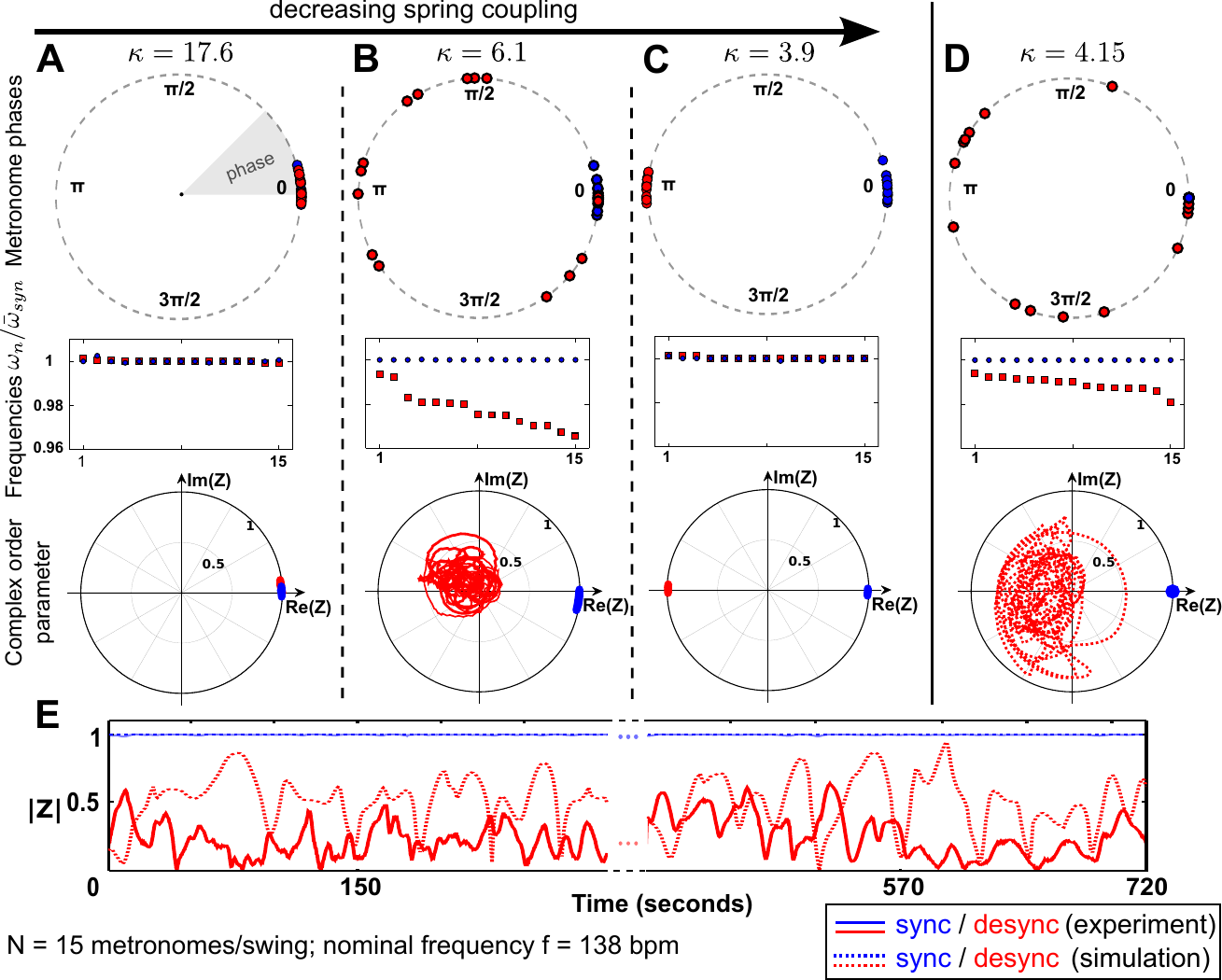}
 \caption{Chimeras emerge with intermediate spring rate $\kappa$ in a 'competition' zone between two fully synchronous modes. With decreasing $\kappa$, we observe a transition from in-phase synchronization  ({\bf A}), over chimeras ({\bf B}) to anti-phase synchronization ({\bf C}). The transition region also exhibits phase-clustered states and partial chimeras. ({\bf D, E}) Simulations share all features of the experimental chimera. Data related to the synchronous and asynchronous population are coded in blue and red, respectively.  Angular frequencies are normalized with the average frequency of the synchronized population $\bar{\omega}_{syn}$.
}
\label{figure2}
\end{figure*}
We start with a fixed frequency and gradually decrease $\kappa$. 
For sufficiently large  $\kappa$, the spring is effectively so stiff that the two swings act like one, and metronomes evolve to a synchronized in-phase motion (IP), so that the complex order parameters overlap with $|Z_{1,2}|\approx1$ (Fig.~\ref{figure2}A and Movie S1).
For low $\kappa$, we observe that the two metronome populations settle into synchronized anti-phase motion (AP), where the order parameters and phases are separated in the complex plane by 180$^\circ$ with $|Z_{1,2}|\approx1$ (Fig.~\ref{figure2}C and Movie S2).
These synchronization modes correspond to the two eigenmodes of the swing-spring system.
For intermediate $\kappa$, however, we observe chimeras (Fig.~\ref{figure2}B and Movie S3). While one of the metronome populations is fully synchronized with  $|Z|\approx1$, the other population is desynchronized. The trajectory of the order parameter of the desynchronized population describes a cloud in the complex plane with $|Z|<1$. The phases of the desynchronized population are spread over the entire interval $[-\pi,\pi]$ and the time-averaged frequencies are non-identical. As we increase $\kappa$, numerical simulations (see below) reveal that this cloud bifurcates off the AP mode, traverses the complex plane and eventually collapses into the stable IP synchronization mode (Fig.~\ref{figure4}B).
None of the metronomes in the  desynchronized population is locked to the synchronized population either, demonstrating truly unlocked motion.  
Chimeras were consistently found for both SD or DS symmetries, ruling out chimeras as a result of asymmetry or pinning due to heterogeneities. Further, chimeras were not transient so that the desynchronized population remained desynchronized, i.e., a DS or SD configuration remained for the entire duration of the experiment, typically lasting up to 1500 oscillation cycles.

\begin{figure*}[htp!]
 \centering
 \includegraphics[width=1\textwidth]{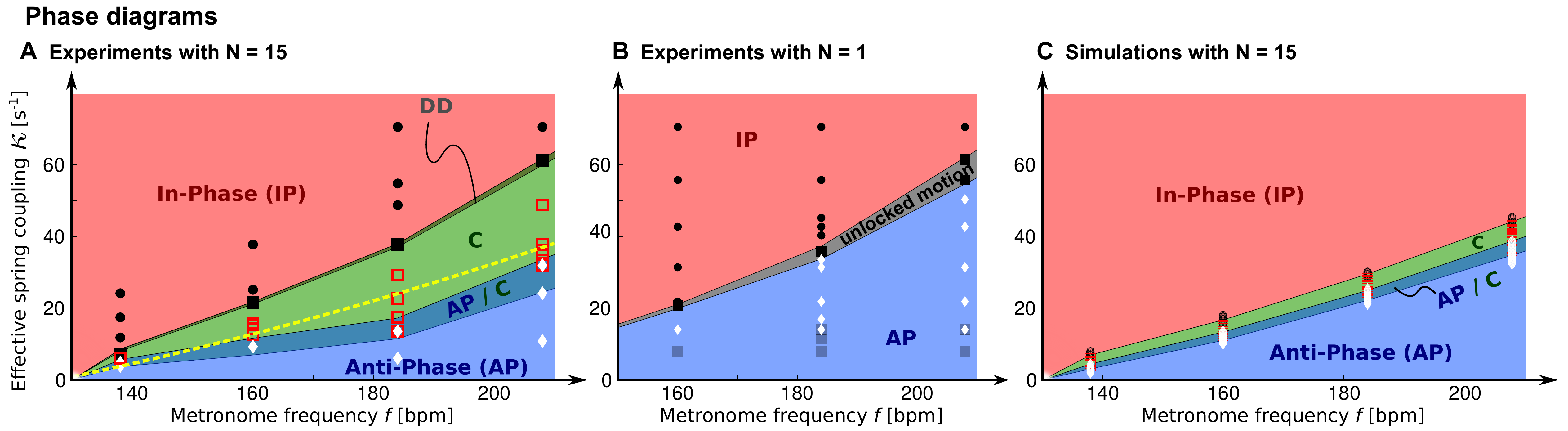}
 \caption{
 Phase diagrams from experiment for  $N=15$ ({\bf A}) and  $N=1$ ({\bf B}) metronome(s) per swing, and from numerical simulations with $N=15$ metronomes ({\bf C}) with metronome frequency $f$ versus effective spring coupling $\kappa = k/M (l/L)^2$. 
 IP (red) and AP (blue) synchronization modes surround the chimera parameter region C (green) and the bistable AP/C region with chimeras and AP synchronization.
 Symbols represent data points (color shadings are guides only).  Region { C}, centered around the resonance curve of the swings' anti-phase mode (dashed) defined by $f\cdot\pi/60 =\sqrt{\Omega^2+2\kappa}$, exhibits chimeras and other partially synchronized states. The bistable region {AP/C} exhibits chimera-like and synchronized anti-phase states; { DD} represents a region where neither population synchronize fully. For $N=1$ we find a similar region of unlocked motion, where the metronomes never synchronize.
 The phase diagram from numerical simulations for identical metronomes exhibits the same qualitative structure as the experiment, except that the width of region C is smaller (see SI Appendix). 
 Parameter space in experiments and simulations was sampled with varying spring coupling $\kappa$ for metronome frequencies $f=$ 138, 160, 184, 208 beats per minute.
 }
\label{figure3}	
\end{figure*} 

Chimeras are sandwiched in a region between AP and IP modes consistently across various metronome frequencies (Fig.~\ref{figure3}A).  
Remarkably, we also find other asynchronous states, including phase clustered states~\cite{GolombHansel1992} (see SI Appendix Fig.~S4), a `partial chimera', where only a fraction of the asynchronous population is frequency-locked (SI), and states with oscillation death~\cite{Pantaleone2002,Bennett2002}. 
Additionally, we observe a region of bi-stability of chimeras and AP synchronized motion. Closer to the edge of the IP region, we find a narrow slice where neither of the metronome populations can achieve synchrony (DD states): even when initialized with SD or DS conditions, the system loses synchrony completely after a transient time.

We have developed a mathematical model (SI) which we simulated to corroborate our experimental findings and to test situations that cannot be achieved experimentally, such as large metronome populations or perfectly identical frequencies. 
The two swings are parametrized by their displacement angles from equilibrium positions,  $\Phi$ and $\Psi$;  the metronome pendula are parametrized by the displacement angles $\phi_i$ and $\psi_i$, respectively.
The metronomes are described as self-sustained oscillators with (harmonic) eigenfrequency $\omega$, damping $\mu_m$ with an amplitude-dependent nonlinearity $D(\phi_i)$ due to the escapement~\cite{Pantaleone2002,Ulrichs2009,Bennett2002}, 
\begin{eqnarray}
\ddot{\phi}_i + \sin{\phi_i}+\mu_m\,{\dot{\phi_i}}\,D(\phi_i) + \frac{\omega^2}{\Omega^2}\cos{\phi}_i\,\ddot{\Phi}=0,\
\end{eqnarray}
where the terms represent (from left to right) pendulum inertia, gravitational force of restitution, damping, and the driving swing inertia, and the dots represent derivatives with respective to time $\tau = \omega t$.
In turn, the swings of length $L$ are described as harmonic oscillators with eigenfrequency $\Omega=\sqrt{g/L}$ and damping $\mu_s$. A swing is driven by the metronomes and the neighboring swing, to which it is coupled with a spring of strength $\kappa$,
\begin{eqnarray}\ddot{\Phi} 
 + \Omega^2 \Phi       
 -\kappa(\Psi - \Phi)     
 +\mu_s\dot{\Phi} 
 + \frac{x_0}{L} \sum_{k=1}^N\partial_{\tau\tau}\sin{\phi_k}
      = 0
 ,
\end{eqnarray}
where terms (from left to right) are swing inertia, force of restitution, spring coupling, friction, and the inertia summed over all metronomes on the same swing. While $\kappa$ determines the inter-population coupling strength, the global coupling strength depends on the ratio of the two eigenfrequencies, $(\omega/\Omega)^2$, and the ratio of the characteristic swing motion amplitude $x_0$ and the swing length $L$. Using conditions similar to our experiments (but without frequency spread), chimeras obtained from simulations (Fig.~\ref{figure2}D, E) and the resulting phase diagram (Fig.~\ref{figure3}C) agree qualitatively very well with experiments (quantitive differences are likely due to the ad-hoc metronome model and potential discrepancies in parametrization, see SI Appendix).
Bi-stability of fully synchronized (SS) and symmetry breaking (SD, DS) states is a hallmark of the chimera instability~\cite{Abrams2008}, which is in distinct contrast to other symmetry breaking scenarios mediated  via supercritical transitions~\cite{Abrams2004}. It is therefore interesting to note that chimera states may coexist with AP synchronization modes in certain regions of the phase diagram (Figs.~\ref{figure3}A and~\ref{figure3}C).

\begin{figure*}[htp!]
 \centering
 \includegraphics[width=1\textwidth]{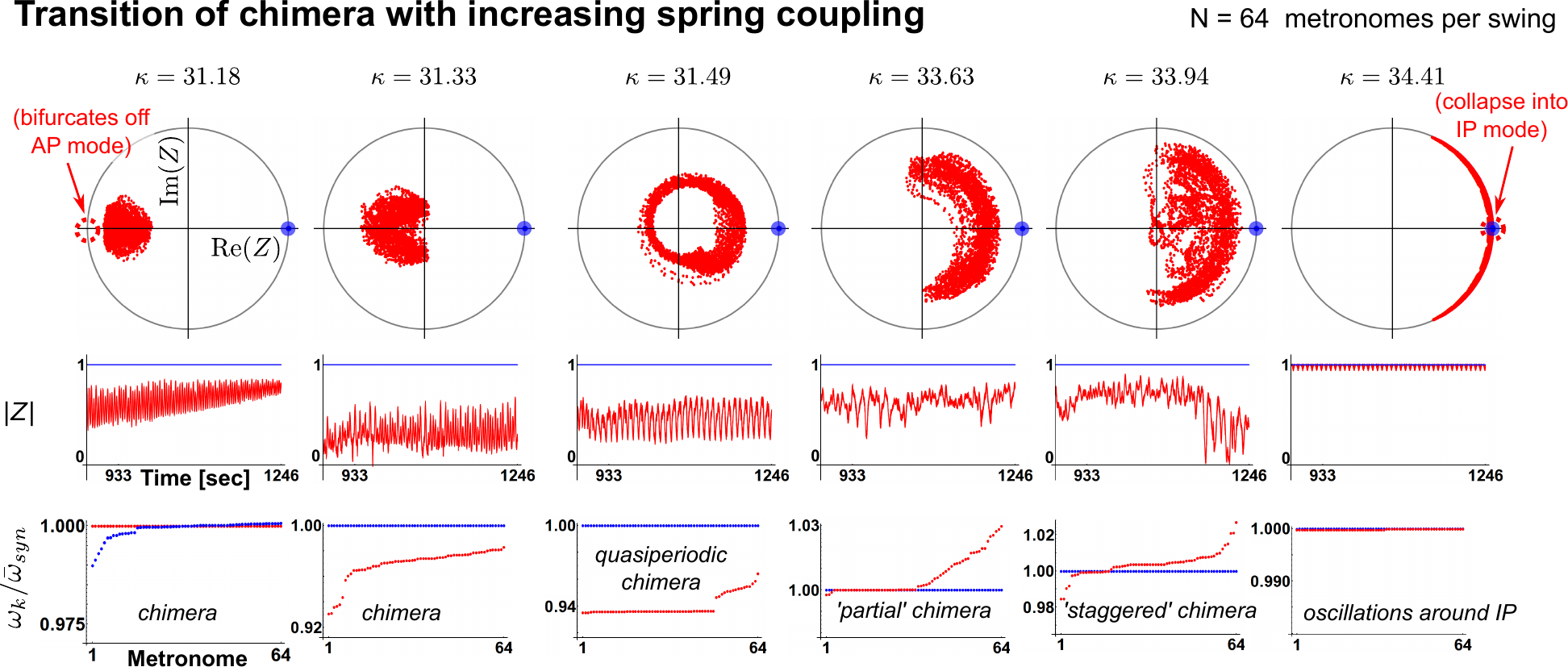}
 \caption{
 Traversal of order parameter cloud with increasing spring coupling $\kappa$. A transition through a rich spectrum of chimera states becomes evident.
 Numerical simulations are carried out with $N=64$ metronomes (see SI Appendix for parameters).
 As $\kappa$ increases, the complex order parameter $Z_p$ bifurcates off from the AP mode at 180$^\circ$ and travels to the right where it snaps into the IP synchronization at 0$^\circ$. 
 The top row displays complex order parameter $Z$; middle row the magnitude $|Z_p|$ and the bottom row angular frequencies, normalized with the average frequency of the synchronized population $\bar{\omega}_{\rm syn}$.
 Synchronized population is shown in blue and desynchronized in red.
 }
 \label{figure4}
\end{figure*}


Notably, when metronomes on each swing synchronize in IP or AP modes, one envisages that the swings -- together with the attached metronomes -- collectively behave like two `giant' metronomes. These modes correspond to excitations of the eigenmodes of the swing pair with frequencies $\Omega$ (IP) and $\sqrt{2\kappa+\Omega^2}$ (AP). 
Indeed, for $N = 1$ metronome per swing (Fig.~\ref{figure1}B), we find that due to momentum transfer, the swing strictly follows the motion of the attached metronome pendulum: the system behaves like Huygens' experiment, i.e. with clocks replaced by metronomes. The metronomes settle into AP and IP synchronization modes for weak and strong coupling $\kappa$, respectively, like in modern reconstructions of Huygens' setup~\cite{Bennett2002}. Additionally, we find a small region where unlocked motion is possible (see Fig.~\ref{figure3}B).

Generalizing Huygens' experiment by adding internal degrees of freedom (i.e., metronomes) on each swing allows for much richer complex dynamics. A rich tapestry of complex states is uncovered (Fig.~\ref{figure4}) in a transition from the AP to IP synchronization as the spring coupling $\kappa$  is increased. In addition to chimeras, these include phase-clustered states~\cite{Tinsley2012}, a `clustered chimera' where oscillators are attracted to a clustered state but cannot quite attain frequency-locking, a partial chimera where the asynchronous population is partially locked, and a quasi-periodic chimera~\cite{Pikovsky2008,Bordyugov2010}. The situation is aptly captured by the notion of ``More is different''~\cite{Anderson1972}: 
additional internal degrees of freedom open a door to unexpected complex behavior, i.e., unanticipated by mere extrapolation of simple collective behavior~\cite{Anderson1972}. 
Using Huygens' term of the ``odd sympathy of clocks''~\cite{Huygens1967ab} to denote synchrony, the observed asymmetric behavior might be described as an ``anti-pathetic sympathy of clocks''.

Chimeras and other partly synchronous states emerge as a competition in an intermediate regime between IP and AP synchronization modes: as a result, both modes are destroyed, so that only one of the `giant' metronomes wins the tug-of-war and remains synchronous, while the other one is broken apart. 
The resulting asymmetry is characterized by the domination of one giant over the other, i.e., the synchronous population forces the  asynchronous population~\cite{Childs2008}, acting like an energy sink.
Remarkably, we find that the parameter region with chimera-like behavior is centered around the resonance curve related to the swings' anti-phase eigenmode (Fig.~\ref{figure3}A): near resonance, the fabric of uniform synchronization is torn.

\section{Discussion}
By devising a mechanical system composed of just two swings, a spring and a number of metronomes~\cite{Pantaleone2002}, we have extended Huygen's original experiment~\cite{Huygens1967ab,Bennett2002} and demonstrated how chimeras emerge in the framework of classical mechanics. 
Recent experiments~\cite{Tinsley2012,Hagerstrom2012} could only produce chimeras by exploiting sophisticated computer-controlled feedback, and the time delay of the coupling had to be carefully crafted in addition to tuning its strength; by contrast, in our realization, chimeras emerge generically using merely a spring, without any need to adjust parameters other than the coupling strength. 
Notably, our setup is composed of basic mechanical elements such as inertia, friction, and spring rate, which have exact or generalized analogues in other areas such as electronic~\cite{Wiesenfeld1998,Temirbayev2012}, optomechanical~\cite{Zhang2012}, chemical~\cite{Kiss2002} and microbial systems or genetic circuits~\cite{Danino2010}. 
The model we propose shows that the complex synchronization patterns found in the experiments are described by elementary dynamical processes that occur in diverse natural and technological settings. This raises the question whether chimeras may have already been observed in such systems, but remained unrecognized as such. 
For instance, our model equations translate directly to recent theoretical studies of synchronization in power grids~\cite{Rohden2012,Dorfler2013,Motter2013} and optomechanical crystals~\cite{Eichenfield2009,Heinrich2011}.  
Consequently, as power grid network topologies evolve to incorporate growing sources of renewable power, the resulting decentralized, hierarchical networks~\cite{Rohden2012} may be threatened by chimera states, which could lead to large scale partial blackouts and unexpected behavior. 
On the other hand, we envision that multistable patterns of synchrony and desynchrony~\cite{Martens2010bistable}
can be exploited to build on-chip memories and computers based on arrays of micro-mechanical devices~\cite{Zhang2012}.
We expect the physical mechanisms that we uncovered here will have important and far-reaching ramifications in the design and usage of such technologies and in understanding chimera states in nature.

\subsection*{Materials and Methods}

\paragraph*{Experiments.}
Two swings are suspended by four light hollow aluminum rods of 50 cm length (outer and inner diameters are 10 mm and 9 mm, respectively). The swings are attached to the rods via low friction ball bearings to ensure smooth motion of the swings. The upper rod ends are attached in the same way on a large rigid support frame. The distance between the support frame and the board is set to ${L=22} $ cm. The motion of the two swings is constrained so that it  to high precision can only occur in the $ (x,y)$-plane. Each swing is made of a $500~$mm x $600~$mm x $1~$mm perforated aluminium plate.  The total weight of each plate is  $ 915$ g $ \pm 4$ g. Each swing is loaded with $ N=15$ metronomes of weight $ 94$ g. The total weight of swing and metronomes is $ M=2.3$ kg. 
Two precision steel springs (Febrotec GmbH; spring constant $ k = 34\,N/m$) are firmly attached with clamps to the two adjacent swing rods (Fig.~\ref{figure1}A) at a distance $ l$ above the pivot point. Adjusting the spring lever $ l$ changes the effective spring strength $ \kappa=k/M (l/L)^2$.
An experiment is started with a careful symmetry check of the system, by ensuring that the initial friction $  \mu_s$ is the same on both swings. The metronome's nominal frequency is set to identical values $ \omega_n$. We then connect the two swings with the spring, firmly set at a distance $ l$ above the pivot points. The motion of the metronomes and the swings is recorded by video recording under UV illumination using a Nikon D90 camera mounted with an $ 18-55~$ mm DX format lens. Each experiment is repeated with inverted roles of the swings (i.e. a DS experiment is followed by an SD experiment), so that the left-right symmetry is checked thoroughly. 

\paragraph*{Simulations.}
Simulations were carried out with identical metronomes until a stationary state was reached (typically, $ \sim~2000$ oscillation cycles). The phase diagram (Fig.~\ref{figure3}C) was obtained by fixing the nominal metronome frequency $ f$ and then gradually increasing the effective spring rate $ \kappa$ (using similar parameters as in the experiment and  $ N=15$ metronomes per swing). For each parameter step, synchronous IP and AP states were continued quasi-adiabatically, whereas simulations resulting in chimera-like states were re-initialized with randomized phases in one of the populations (see SI Appendix).


\begin{acknowledgments}
We thank H.~Stone, S.~Strogatz, K.~Showalter, A.~Pikovsky, M.~Rosenblum, and S.~Herminghaus for useful comments, and H.~J. Martens and U.~Krafft for advice on the experimental design.
We dedicate this paper in the fond memory of U.~Krafft. This work was partly supported by the Human Frontier Science Program (to S.T.).
\end{acknowledgments}

\bibliographystyle{unsrt}

\begin{thebibliography}{10}

\bibitem{Huygens1967ab}
C.~ Huygens.
\newblock {\em {{\OE}uvres compl{\`e}tes, Vol. 15}}.
\newblock Swets \& Zeitlinger B. V., Amsterdam, 1967.

\bibitem{Michaels1987}
D.~C. Michaels, E.~P. Matyas, and J.~Jalife.
\newblock {Mechanisms of sinoatrial pacemaker synchronization: a new
  hypothesis}.
\newblock {\em Circ. Res.}, 61(5):704--714, November 1987.

\bibitem{Buck1968}
J.~Buck and E.~Buck.
\newblock {Mechanism of Rhythmic Synchronous Flashing of Fireflies: Fireflies
  of Southeast Asia may use anticipatory time-measuring in synchronizing their
  flashing}.
\newblock {\em Science}, 159(3821):1319--1327, March 1968.

\bibitem{Strogatz2005}
S.~H. Strogatz, D.~M. Abrams, A. McRobie,  B. Eckhardt, and E.  Ott.
\newblock {Theoretical Mechanics: Crowd Synchrony on the Millennium Bridge.}
\newblock {\em Nature}, 438(7064):43--4, November 2005.

\bibitem{Liu1997}
C. Liu, D.~R. Weaver, S.~H. Strogatz, and S.~M. Reppert.
\newblock {Cellular Construction of a Circadian Clock: Period Determination in
  the Suprachiasmatic Nuclei}.
\newblock {\em Cell}, 91(6):855--860, December 1997.

\bibitem{Wiesenfeld1998}
K. Wiesenfeld, P. Colet, and S. Strogatz.
\newblock {Frequency locking in Josephson arrays: Connection with the Kuramoto
  model}.
\newblock {\em Physical Review E}, 57(2):1563--1569, February 1998.

\bibitem{Kiss2002}
I.~Z Kiss, Y. Zhai, and J.~L. Hudson.
\newblock {Emerging coherence in a population of chemical oscillators.}
\newblock {\em Science (New York, N.Y.)}, 296(5573):1676--8, May 2002.

\bibitem{Taylor2009}
A.~F. Taylor, M.~R. Tinsley, F. Wang, Z. Huang, and K. Showalter.
\newblock {Dynamical quorum sensing and synchronization in large populations of
  chemical oscillators.}
\newblock {\em Science (New York, N.Y.)}, 323(5914):614--7, January 2009.

\bibitem{Dano1999}
S. Dano, P~G. Sorensen, and F. Hynne.
\newblock {Sustained oscillations in living cells.}
\newblock {\em Nature}, 402(6759):320--2, November 1999.

\bibitem{Massie2010}
T.~M. Massie, B. Blasius, G. Weithoff, U. Gaedke, and G.~F. Fussmann.
\newblock {Cycles, phase synchronization, and entrainment in single-species
  phytoplankton populations.}
\newblock {\em Proc. Natl. Acad. Sci.}, 107(9):4236--41, March 2010.

\bibitem{Motter2010a}
A.~E. Motter.
\newblock {Spontaneous Synchrony Breaking}.
\newblock {\em Nature Physics (News and Views)}, 6(3):164--165, 2010.

\bibitem{Kuramoto2002}
Y.~Kuramoto and D.~Battogtokh.
\newblock {Coexistence of Coherence and Incoherence in Nonlocally Coupled Phase
  Oscillators}.
\newblock {\em Nonlinear Phenomena in Complex Systems}, 4:380 -- 385, 2002.

\bibitem{Abrams2004}
D. Abrams and S.~H. Strogatz.
\newblock {Chimera States for Coupled Oscillators}.
\newblock {\em Physical Review Letters}, 93(17):174102, October 2004.

\bibitem{Abrams2008}
D.~M. Abrams, R.~Mirollo, S.~H. Strogatz, and D.~A. Wiley.
\newblock {Solvable model for chimera states of coupled oscillators}.
\newblock {\em Phys. Rev. Lett}, 101:084103, 2008.

\bibitem{Montbrio2004}
E. Montbri\'{o}, J. Kurths, and B. Blasius.
\newblock {Synchronization of two interacting populations of oscillators}.
\newblock {\em Physical Review E}, 70(5):056125, November 2004.

\bibitem{Omelchenko2008}
O.~E. Omel’chenko, Y.~L Maistrenko, and P.~A. Tass.
\newblock {Chimera states: The natural link between coherence and incoherence}.
\newblock {\em Physical Review Letters}, 100(4):044105, 2008.

\bibitem{Pikovsky2008}
A.~Pikovsky and M.~Rosenblum.
\newblock {Partially integrable dynamics of hierarchical populations of coupled
  oscillators}.
\newblock {\em Phys. Rev. Lett}, 101:264103, 2008.

\bibitem{Bordyugov2010}
G. Bordyugov, A. Pikovsky, and M. Rosenblum.
\newblock {Self-Emerging and Turbulent Chimeras in Oscillator Chains}.
\newblock {\em Phys. Rev. E}, 36:035205, 2010.

\bibitem{Martens2010bistable}
E.~A. Martens.
\newblock {Bistable Chimera Attractors on a Triangular Network of Oscillator
  Populations}.
\newblock {\em Physical Review E}, 82(1):016216, July 2010.

\bibitem{Martensswc2010}
E.~A. Martens, C.~R. Laing, and S.~H. Strogatz.
\newblock {Solvable Model of Spiral Wave Chimeras}.
\newblock {\em Physical Review Letters}, 104(4):044101, January 2010.

\bibitem{Laing2009}
C.~R. Laing.
\newblock {The dynamics of chimera states in heterogeneous Kuramoto networks}.
\newblock {\em Physica D: Nonlinear Phenomena}, 238(16):1569--1588, August
  2009.

\bibitem{Laing2012a}
C.~R. Laing, K. Rajendran, and I.~G. Kevrekidis.
\newblock {Chimeras in random non-complete networks of phase oscillators}.
\newblock {\em Chaos: An Interdisciplinary Journal of Nonlinear Science},
  22(1):013132, 2012.
  
\bibitem{Olmi2010}
S.~Olmi, A.~Politi, and A.~Torcini.
\newblock{Collective chaos in pulse-coupled neural networks}.
\newblock {Europhysics Letters}, 92, 60007, 2010.


\bibitem{Phillips1993}
J.~R. Phillips, H.~S.~J. Van der Zant, J. White, and T.~P.Orlando.
\newblock {Influence of induced magnetic fields on the static properties of
  Josephson-junction arrays}.
\newblock {\em Physical Review B}, 47(9), 1993.

\bibitem{Swindale1980}
N.~V. Swindale.
\newblock {A Model for the Formation of Ocular Dominance Stripes}.
\newblock {\em Proc. R. Soc. London B}, 208:243--264, 1980.

\bibitem{Tinsley2012}
M.~R. Tinsley, S. Nkomo, and K. Showalter.
\newblock {Chimera and phase-cluster states in populations of coupled chemical
  oscillators}.
\newblock {\em Nature Physics}, 8(8):662--665, July 2012.

\bibitem{Hagerstrom2012}
A.~M. Hagerstrom, T.~E. Murphy, R. Roy, P. H\"{o}vel, I.  Omelchenko, and E. Sch\"{o}ll.
\newblock {Experimental observation of chimeras in coupled-map lattices}.
\newblock {\em Nature Physics}, 8(8):658--661, July 2012.

\bibitem{Pantaleone2002}
J. Pantaleone.
\newblock {Synchronization of metronomes}.
\newblock {\em American Journal of Physics}, 70(10):992, 2002.

\bibitem{Ulrichs2009}
H. Ulrichs, A. Mann, and U. Parlitz.
\newblock {Synchronization and chaotic dynamics of coupled mechanical
  metronomes.}
\newblock {\em Chaos (Woodbury, N.Y.)}, 19(4):043120, December 2009.

\bibitem{GolombHansel1992}
D. Golomb, D. Hansel, B. Shraiman, and H. Sompolinsky.
\newblock {Clustering in globally coupled phase oscillators}.
\newblock {\em Physical Review A}, 45(6):3516--3531, 1992.

\bibitem{Bennett2002}
M. Bennett, M. Schatz, and K. Wiesenfeld.
\newblock {Huygen's clocks}.
\newblock {\em Proc. R. Soc. London A}, 458(2019):563--579, 2002.

\bibitem{Anderson1972}
P.~W. Anderson.
\newblock {More is different}.
\newblock {\em Science}, 177(4047):393--396, 1972.

\bibitem{Childs2008}
L.~M. Childs and S.~H. Strogatz.
\newblock {Stability diagram for the forced Kuramoto model.}
\newblock {\em Chaos (Woodbury, N.Y.)}, 18(4):043128, December 2008.

\bibitem{Temirbayev2012}
A. Temirbayev, Z. Zhanabaev, S. Tarasov, V.  Ponomarenko, and M. Rosenblum.
\newblock {Experiments on oscillator ensembles with global nonlinear coupling}.
\newblock {\em Physical Review E}, 85(1):015204(R), January 2012.

\bibitem{Zhang2012}
M. Zhang, G. Wiederhecker, S. Manipatruni, A. Barnard, P.  Mceuen, and M. Lipson.
\newblock {Synchronization of Micromechanical Oscillators Using Light}.
\newblock {\em Phys. Rev. Lett}, 109, 2012.

\bibitem{Danino2010}
T. Danino, O. Mondrag\'{o}n-Palomino, L. Tsimring, and J. Hasty.
\newblock {A synchronized quorum of genetic clocks.}
\newblock {\em Nature}, 463(7279):326--30, January 2010.

\bibitem{Rohden2012}
M. Rohden, A. Sorge, M. Timme, and D. Witthaut.
\newblock {Self-Organized Synchronization in Decentralized Power Grids}.
\newblock {\em Physical Review Letters}, 109(6):064101, August 2012.

\bibitem{Dorfler2013}
F. D\"{o}rfler, M. Chertkov, and F. Bullo.
\newblock {Synchronization in complex oscillator networks and smart grids.}
\newblock {\em Proceedings of the National Academy of Sciences of the United
  States of America}, 110(6):2005--10, February 2013.

\bibitem{Motter2013}
A.~E. Motter, S.~A. Myers, M. Anghel, and T. Nishikawa.
\newblock {Spontaneous synchrony in power-grid networks}.
\newblock {\em Nature Physics}, Advance on, 2013.

\bibitem{Eichenfield2009}
M. Eichenfield, J. Chan, R.~M. Camacho, K.~J. Vahala, and O. Painter.
\newblock {Optomechanical crystals.}
\newblock {\em Nature}, 462(7269):78--82, November 2009.

\bibitem{Heinrich2011}
G.~H. Heinrich, M. Ludwig, J. Qian, B. Kubala, and F.  Marquardt.
\newblock {Collective Dynamics in Optomechanical Arrays}.
\newblock {\em Physical Review Letters}, 107(4):8--11, July 2011.


\end{thebibliography}

\end{document}